\documentclass[preprint]{ptephy_v1}
\usepackage{amssymb}
\usepackage{graphicx}
\numberwithin{equation}{section}

\newcommand{\dl}{\delta}
\newcommand{\cb}{\bar{c}}
\newcommand{\lb}[1]{\label{#1}}
\newcommand{\cl}{\mathcal{L}}
\newcommand{\bb}{\bar{B}}

\newcommand{\vp}{\varphi}
\newcommand{\ptl}{\partial}

\newcommand{\ct}{\tilde{C}}
\newcommand{\cvp}{\tilde{\varphi}}
\newcommand{\ph}{\mathrm{phys}}

\begin{document}

\title{Condensate $\langle A_{\mu}^+A_{\mu}^-\rangle$ 
and massive magnetic potential in Euclidean gauge theories
}

\author{\name{\fname{Hirohumi} \surname{Sawayanagi}}{1}}

\address{\affil{1}{National Institute of Technology, Kushiro College, Kushiro, 084-0916, Japan}
\email{sawa@kushiro-ct.ac.jp}}


\begin{abstract}
Euclidean $SU(2)$ gauge theory is studied in a nonlinear gauge.  
In this theory, ghost condensation happens and gauge fields acquire tachyonic masses.  
It is shown that these tachyonic masses are removed by a gauge field condensate 
$\langle A_{\mu}^+A_{\mu}^-\rangle$.  Because of the ghost condensation, monopole 
solutions are included naturally.  We find the condensate $\langle A_{\mu}^+A_{\mu}^-\rangle$ 
makes the magnetic potential massive.  
\end{abstract}

\subjectindex{B0,B3,B6}

\maketitle

\section{Introduction}

     In non-Abelian gauge theories, a magnetic monopole may play an important role.  In the 
dual superconductor model of color confinement \cite{rip}, monopole condensation 
is expected.  As a result, a dual gauge field \cite{zwa} acquires a mass, and 
color confinement may happen.  

     How can monopoles appear without Higgs fields?  One way is the Abelian projection 
by 't~Hooft \cite{tH1}.  Let us consider a field $X(x)$ which belongs to 
the adjoint representation, and perform the Abelian gauge fixing.  
When we diagonalize $X(x)$, there may be points that the eigenvalues of $X(x)$ degenerate.  
Monopoles appear at these points.  Another way is to introduce a unit color vector $\hat{n}^A(x)$ 
in the internal space \cite{cho,kon}.  Non-Abelian magnetic potential is defined 
by 
\begin{equation}
    C_{\mu}^A =  \frac{-1}{g}(\hat{n}\times \ptl_{\mu}\hat{n})^A,  \lb{101}
\end{equation}
and the gauge field $A_{\mu}^A(x)$ is decomposed 
by using $\hat{n}^A$.  This model is called the extended QCD \cite{cho1}.  

     What is the origin of $\hat{n}^A$?  Is it possible to make the magnetic potential massive?  
In this paper, we propose a scenario which produces $\hat{n}^A$ and massive 
$C_{\mu}^A$.  For simplicity, we consider the SU(2) gauge theory in Euclidean space-time, and sometimes call the 
gauge field gluon.  
In the next section, we briefly review ghost condensation and 
tachyonic gluon masses.  In Sect. 3, it is shown that these tachyonic masses are 
removed by the vacuum expectation value (VEV) $\langle A_{\mu}^+A_{\mu}^-\rangle$.  
Under the ghost condensation, we can include a magnetic monopole as a classical solution.  
In Sect. 4, we introduce it in the Abelian gauge.  
In the presence of the VEV $\langle A_{\mu}^+A_{\mu}^-\rangle$, 
a magnetic potential is expected to become massive.   By using the background covariant gauge, 
we confirm this expectation in Sect. 5.  In Sect. 6, to remove the Dirac string, we perform 
a singular gauge transformation, and derive the extended QCD with massive 
$C_{\mu}^A$.  
In Sect. 7, the case of three dimensional space-time is discussed briefly.  
Section 8 is devoted to a summary and comments.  In Appendix A, 
it is shown that $\langle (A_{\mu}^3)^2\rangle$ vanishes.  
The tachyonic gluon mass is derived in the 
background covariant gauge in Appendix B.  The singular gauge transformation used in Sect. 6 is 
summarized in Appendix C.  The BRS symmetry and the breakdown of the global gauge symmetry are 
explained in Appendix D.

\section{Ghost condensation and tachyonic gluon mass}

       We consider the SU(2) gauge theory with structure constants $f_{ABC}$.  
Using the notations 
\[ F\cdot G=F^AG^A, \quad (F\times )^{AB}=f_{ACB}F^C, \quad
(F\times G)^A=f_{ABC}F^BG^C, \quad 
A=1,2,3, \]
the Lagrangian in the nonlinear gauge is given by ~\cite{hs1}
\begin{align}
 \cl &= \cl_{inv}+ \cl_{NL},\quad \cl_{inv}=\frac{1}{4}F_{\mu\nu}^2, \notag \\
  \cl_{NL} &=  B\cdot \partial_{\mu}A_{\mu}+i\cb\cdot\partial_{\mu}D_{\mu}c-
 \frac{\alpha_1}{2}B^2-\frac{\alpha_2}{2}\bb^2 -B\cdot w,   \lb{201}
\end{align}
where $\bb = -B+ ig\cb \times c$, $\alpha_1$ and $\alpha_2$ are gauge parameters, and 
$w$ is a constant.  
Introducing the auxiliary field $\varphi$, which represents $\alpha_2 \bb$, $\cl_{NL}$ 
is rewritten as 
\begin{equation}
 \cl_{\vp}=-\frac{\alpha_1}{2}B^2 
 +B\cdot (\ptl_{\mu}A_{\mu}+\vp -w)+i\cb \cdot(\ptl_{\mu}D_{\mu}+g\vp \times )c 
 +\frac{\vp^2}{2\alpha_2}.  \lb{202}
\end{equation}
In Ref.~\cite{hs2}, we have shown that $g \vp$ acquires the VEV 
$v=g\vp_0$ under an energy scale $\mu_0$.  
Dividing the quantum fluctuation $\cvp(x)$ from $\vp_0$, we substitute $\vp(x)=\vp_0 + \cvp(x)$ into Eq.(\ref{202}), 
and choose the constant $w=\vp_0$.  This choice is necessary to maintain the BRS symmetry \cite{kug}.  

Then Eq.(\ref{202}) becomes 
\begin{equation}
 \cl_{\vp}'=-\frac{\alpha_1}{2}B^2 
 +B\cdot (\ptl_{\mu}A_{\mu}+\cvp )+i\cb \cdot(\ptl_{\mu}D_{\mu}+g\cvp \times + g\vp_0 \times )c .
   \lb{203}
\end{equation}
Because of the VEV $\vp_0$, the global SU(2) symmetry breaks down to U(1).  
The relation between this breaking and BRS invariant Green functions is 
explained in Ref.~\cite{hs3}. 
\footnote{In Appendix D, the necessity of $w$ is explained.  The BRS symmetry and the broken global gauge symmetry
are also discussed.} 
 
     Under the ghost condensation, the ghost-loop contribution to the gluon propagator was 
calculated.  When gluon momentum becomes small, 
it is found that the gluon acquires tachyonic mass \cite{hs1,dv}.  
The term $g\vp_0\times$ in the ghost propagator $(\ptl_{\mu}^2+g\vp_0\times)^{-1}$ gives rise to the 
tachyonic mass.  
Now we choose the VEV in the third direction as $\langle \vp^A\rangle=\vp_0\delta^{A3}$, 
and use the notation $(A_{\mu}^a)^2=(A_{\mu}^1)^2+(A_{\mu}^2)^2$.  
Then, the scale $\mu_0$, the value of the VEV $v=g\vp_0$ and the tachyonic gluon mass term 
in the 4-dimensional Euclidean space (D=4) are summarized as 
\begin{equation}
 \mu_0=\Lambda e^{-4\pi^2/(\alpha_2g^2)},\ v=\left\{\frac{\mu_0^4-\mu^4}{1-e^{-16\pi^2/(\alpha_2g^2)}}\right\}^{1/2}, 
       \ \frac{1}{2}\left(\frac{-g^2v}{64\pi}\right)\left\{(A_{\mu}^a)^2+2(A_{\mu}^3)^2\right\},     \lb{204}
\end{equation}
where $\Lambda$ is a momentum cut-off, and $\mu$ is the momentum scale satisfying $\mu<\mu_0$.  The coupling 
constant $g$ and the gauge parameter $\alpha_2$ in Eq.(\ref{204}) are the quantities at the scale $\Lambda$.  
In the same way, the corresponding quantities in the 3-dimensional Euclidean space (D=3) are 
\begin{equation}
\mu_0=\frac{\alpha_2g_3^2\Lambda}{ \pi^2\Lambda+\alpha_2g_3^2},\ v_3=\frac{2}{16\pi^2}(\alpha_2g_3^2)^2, 
 \ \frac{1}{2}\left(\frac{-\alpha_2}{24\pi^2}g_3^4\right)\left\{(A_{\mu}^a)^2+\frac{3}{2}(A_{\mu}^3)^2\right\},
 \lb{205}
\end{equation}
where $g_3$ and $v_3$ are the coupling constant and the VEV $g\vp_0$ in D=3, respectively.

\section{Condensate $\langle A_{\mu}^+A_{\mu}^-\rangle$}

     From now on, we concentrate on the D=4 case, and try to remove the tachyonic gluon mass.  We write Eq.(\ref{204}) as 
\begin{equation}
 -m_4^2 \left[A_{\mu}^+A_{\mu}^- + \frac{\kappa_4}{2} (A_{\mu}^3)^2\right],\ m_4^2=\frac{g^2v}{64\pi}, \  \kappa_4=2  
 \lb{301}
\end{equation}
where $A_{\mu}^{\pm}=(A_{\mu}^1\pm iA_{\mu}^2)/\sqrt{2}$.

\subsection{The effective potential for $\langle A_{\mu}^+A_{\mu}^-\rangle$}

     When the tachyonic mass term (\ref{301}) exist, the gluon propagator blows up as $p^2\to m_4^2$.  
To avoid it, we return to the Lagrangian $\cl_{inv}+\cl_{\vp}'$, and introduce the source term 
$KA_{\mu}^+A_{\mu}^-$ for the local composite operator (LCO) 
$A_{\mu}^+(x)A_{\mu}^-(x)$. 
\footnote{In this subsection, we write $\hslash$ explicitely.  
For scalar fields $\phi(x)$, the effective potential for the LCO $\phi^2$ is studied in Ref.~\cite{hk}.} 
The partition function is 
\[
 Z(J,K) = \int D\mu \exp \left[ -\frac{1}{\hslash}\left\{S+\int dx J\Psi+ \int dx K A_{\mu}^+A_{\mu}^- \right\} \right], 
\]
where usual source terms are 
\[ J\Psi=\cb \eta + \bar{\eta}c + A_{\mu}^3J_{\mu}^3 + A_{\mu}^+J_{\mu}^- + A_{\mu}^-J_{\mu}^+.  \]
Dividing the action $S$ into the free part and the interaction part as $S=S_{free}+S_{int}$, and performing 
the Gaussian integration, we obtain 
\[
  Z(K) = \exp[-\frac{1}{\hslash}W(K)],\quad W(K)=W_1(K)+W_2(K)+\cdots , 
\]
where $W_n$ is $O(\hslash^n)$, and we set $J=0$ after constructing $W(J,K)$.  
From $Z(K)$, we find 
\begin{equation}
     \frac{\delta W}{\delta K(x)}= \langle A_{\mu}^+(x)A_{\mu}^-(x) \rangle =\hslash \Phi_4(x),  \lb{302}
\end{equation}
and we define the effective action as 
\begin{equation}
   \Gamma(\Phi_4)= W(K)- \hslash \int dx K\Phi_4.   \lb{303}
\end{equation}

     If we write the $A_{\mu}^{\pm}$-related part in $S_{free}$ as 
$-A_{\mu}^+\Delta_{\mu\nu}A_{\nu}^-$, the $K$-dependent term in $W_1$ becomes 
\[ W_1(K)= \hslash \mathrm{Tr}\ln (-\Delta + K).  \]
Thus, at $O(\hslash)$, Eq.(\ref{302}) gives 
\begin{equation}
     \Phi_4(x) = \langle x|\left(-\Delta + K \right)^{-1}_{\mu\mu}|x \rangle.  \lb{304}
\end{equation}

\begin{figure}
\begin{center}
\includegraphics{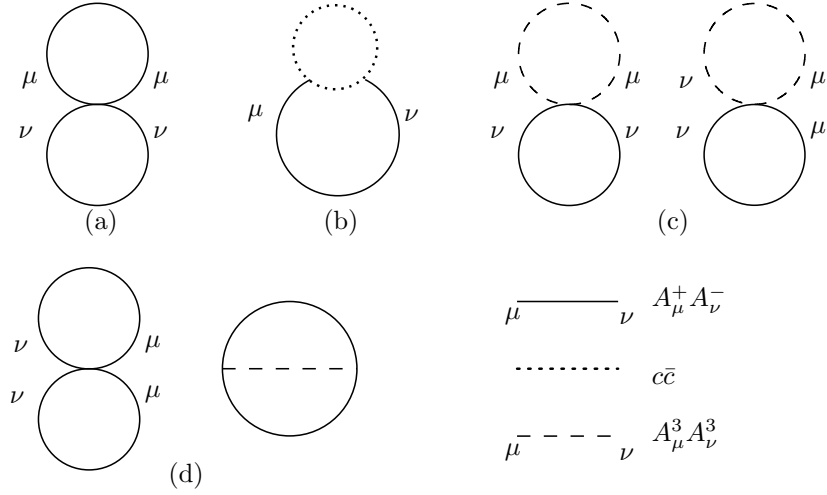}
\caption{The diagrams that contribute to the effective potential $\Gamma(\Phi_4)$ at $O(\hslash^2)$.  
The solid line represents $A_{\mu}^{+}A_{\nu}^{-}$, the dotted line is $c \cb$, 
and the dashed line represents $A_{\mu}^3A_{\nu}^3$.  }
\label{fig1}
\end{center}
\end{figure}

     From $S_{int}$, the diagrams depicted in Fig.1 contribute to $W_2$.  Since the kinetic term $(F_{\mu\nu}^A)^2/4$ 
contains the terms 
\begin{eqnarray}
 \frac{g^2}{4}(f^{ABC}A_{\mu}^BA_{\nu}^C)^2&=&
\frac{g^2}{2}(A_{\mu}^+A_{\mu}^-)^2
+g^2(A_{\mu}^+A_{\mu}^-)(A_{\nu}^3)^2 \nonumber \\
& & - \frac{g^2}{2}(A_{\mu}^+)^2(A_{\nu}^-)^2
-g^2(A_{\mu}^+A_{\mu}^3)(A_{\nu}^-A_{\nu}^3),  \lb{305}
\end{eqnarray}
the diagram Fig.1(a) comes from the vertex $\displaystyle \frac{g^2}{2}(A_{\mu}^+A_{\mu}^-)^2$ 
in Eq.(\ref{305}).  Applying Eq.(\ref{304}), it gives 
$\displaystyle \hslash^2\frac{g^2}{2}\Phi_4^2$.       
The ghost loop in Fig.1(b) yields the term $-m_4^2 \hslash \delta_{\mu\nu}$, which makes $A_{\mu}^{\pm}$ tachyonic.  
Using Eq.(\ref{304}) again, Fig.1(b) gives $-m_4^2 \hslash^2 \Phi_4$.  
The diagrams in Fig.1(c) contain the loops with $A_{\mu}^3$.  In the dimensional regularization, 
as $\displaystyle \int d^4k (1/k^2)=0$, these diagrams vanish.   
If the source $K$ is constant, it plays the role of mass squared $M_4^2$ for the $A_{\mu}^{\pm}$.  So the diagrams in Fig.1(d) can produce 
the term $z_1K\hslash^2\Phi_4$, where $z_1$ is a divergent constant.  This divergence should 
be removed by the renormalization of $K$.  
By summing up these results, $W_2$ becomes $\Gamma_2(\Phi_4)=\hslash^2V(\Phi_4)$, where  
\begin{equation}
 V(\Phi_4)=\frac{g^2}{2}\Phi_4^2-m_4^2\Phi_4.  \lb{306} 
\end{equation}

     Thus, up to $O(\hslash^2)$, we obtain 
\begin{equation}
   \Gamma(\Phi_4)= W_1(K) + \hslash^2 V(\Phi_4)- \hslash \int dx K\Phi_4. \lb{307}   
\end{equation}
This $\Gamma$ satisfies 
\begin{equation}
  \frac{\delta \Gamma}{\delta \Phi_4}= \left(\frac{\delta W_1}{\delta K}-\hslash \Phi_4\right) \frac{\delta K}{\delta \Phi_4}
+ \hslash^2 \frac{d V}{d \Phi_4}-\hslash K=\hslash^2 \frac{d V}{d \Phi_4}-\hslash K,  \lb{308}
\end{equation}
where Eq.(\ref{304}) has been used.  Since $\Gamma$ satisfies $\delta \Gamma/\delta \Phi_4=-\hslash K$, 
from Eq.(\ref{308}), we obtain 
$dV/d\Phi_4=0$.

\begin{figure}
\begin{center}
\includegraphics{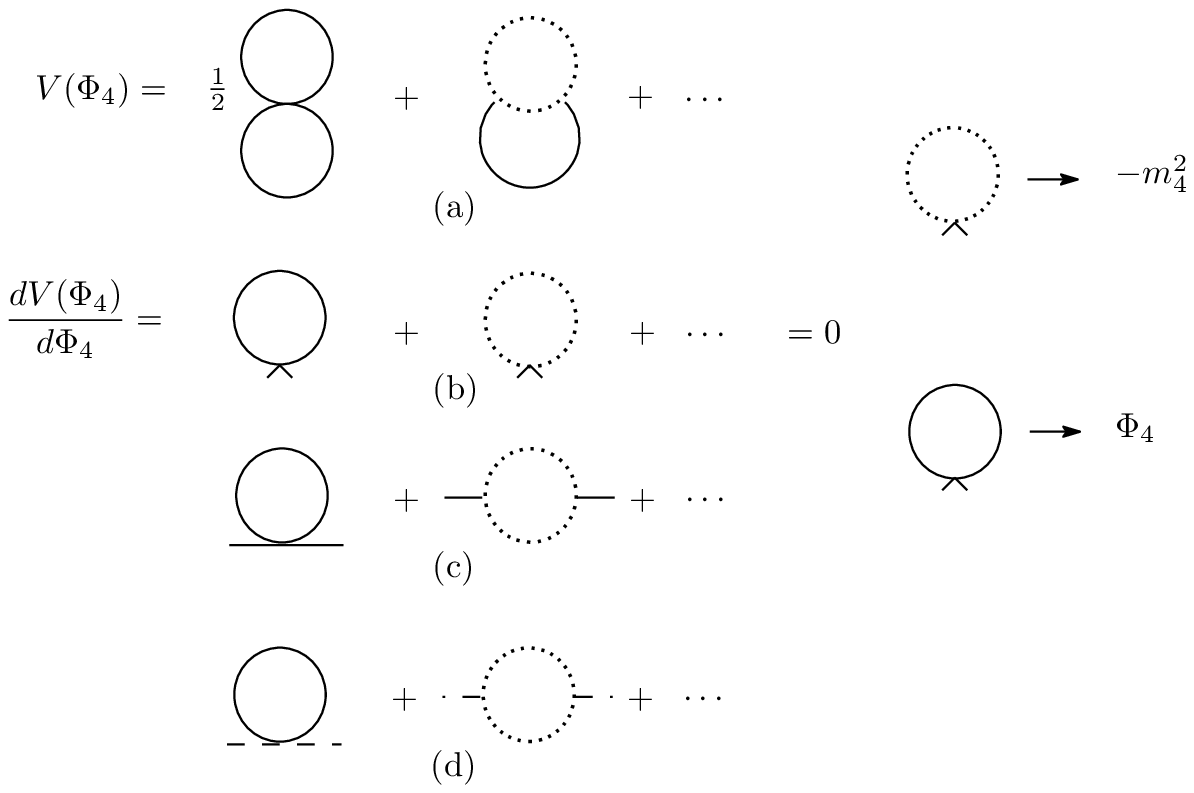}
\caption{The diagrams that contribute to $V(\Phi)$, $dV/d\Phi_4=0$, and 
the mass terms for $A_{\mu}^{\pm}$ and $A_{\mu}^{3}$.}
\label{fig2}
\end{center}
\end{figure}     

     In Fig.2(a), the diagrams which contribute to $V(\Phi_4)$ are depicted.  The closed loop with solid line 
represents $\Phi_4(x)$.  
As we show in Fig.2(b), by removing one closed loop, the diagrams which correspond to $dV/d \Phi_4$ are obtained.  
If we cut one closed loop as in Fig.2(c), the diagrams which contribute to the mass term for 
$A_{\mu}^+(x)A_{\mu}^-(x)$ are obtained.  Therefore, when the condition $dV/d\Phi_4=0$ holds, the two diagrams 
in Fig.2(c) cancel out.  
In the same way, in D=4, the diagrams depicted in Fig.1(d) cancel out.  Namely, the diagrams in Fig.2(c) and Fig.2(d) 
don't contribute to the masses for $A_{\mu}^{\pm}$ and $A_{\mu}^3$, respectively.  

     We summarize the results.  Up to $O(\hslash^2)$, the effective potential $\Gamma$ is (\ref{307}).  As it 
satisfies $\delta \Gamma/\delta \Phi_4=-\hslash K$, we obtain Eq.(\ref{304}) and $dV/d\Phi_4=0$.  
Using Eq.(\ref{306}), $dV/d\Phi_4=0$ gives 
\begin{equation}
 \Phi_4=\langle A_{\mu}^+A_{\mu}^- \rangle=\frac{m_4^2}{g^2}=\frac{v}{64\pi},  \lb{309}
\end{equation}
and Eqs.(\ref{304}) and (\ref{309}) gives 
\begin{equation}
     \frac{m_4^2}{g^2}= \langle x|\left(-\Delta+M_4^2 \right)^{-1}_{\mu\mu}|x \rangle,  \lb{310}
\end{equation}
where we set $K=M_4^2$.  Thus, when $m_4\neq 0$, the condensate $\Phi_4$ exists, and the mass $M_4$ for 
$A_{\mu}^{\pm}$ is determined by Eq.(\ref{310}).  

     As $\Delta_{\mu\nu}$ contains a gauge parameter, Eq.(\ref{310}) is gauge-dependent 
and it requires regularization.  
So, although Eq.(\ref{310}) implies $M_4\neq 0$, the determination of it must be done 
carefully.  We don't try to determine $M_4$ in this paper, and make it a future task.

\subsection{Expedient procedure}

    The above results are obtained simply by the following procedure.  Add the tachyonic mass term (\ref{301}) 
to the Lagrangian.  Then replace $A_{\mu}^+A_{\mu}^-$ as 
\begin{equation}
  A_{\mu}^+A_{\mu}^- \to \Phi_4 + A_{\mu}^+A_{\mu}^-, \quad 
\Phi_4=\langle A_{\mu}^+A_{\mu}^- \rangle,  \lb{311}
\end{equation}
and add the source term $M_4^2 A_{\mu}^+A_{\mu}^-$.  Then we obtain the following terms that contain 
$m_4^2, M_4^2$ or $\Phi_4$:  
\begin{equation}
     V(\Phi_4) + (g^2\Phi_4 -m_4^2)[A_{\mu}^+A_{\mu}^- + (A_{\mu}^3)^2]+M_4^2 A_{\mu}^+A_{\mu}^-,  \lb{312}
\end{equation}
where Eq.(\ref{305}) has been used.  From $dV/d\Phi_4=0$, Eq.(\ref{309}) holds and  Eq.(\ref{312}) 
becomes $M_4^2 A_{\mu}^+A_{\mu}^-$.  Namely, 
the tachyonic masses are removed and ,whereas $A_{\mu}^3$ is massless, $A_{\mu}^{\pm}$ acquire the mass $M_4$.  
In the following sections, we apply this procedure for simplicity.  

     We note, although the component $A_{\mu}^3$ has the tachyonic mass 
in Eq.(\ref{301}) as well, we did not take the source term for $(A_{\mu}^3)^2$ and 
the VEV $\langle (A_{\mu}^3)^2\rangle$ into account.  
In Appendix A, it is shown the VEV $\langle (A_{\mu}^3)^2\rangle$ vanishes.

\section{Inclusion of monopoles}

     The Lagrangian $\cl_{\vp}$ is invariant under the BRS transformation $\delta_B$, 
if the field $\vp$ transforms as $\dl_B \vp=g\vp \times c$.  That is, $\vp$ behaves like a Higgs field in the adjoint 
representation.  
As $\vp$ acquires the VEV $\langle \vp^A\rangle =(v/g)\delta^{A3}$, we can introduce Abelian monopoles in the $A=3$ 
direction \cite{afg}.  Let us consider the classical field $b_{\mu}^A=\ct_{\mu}\delta^{A3}$, which satisfies the 
equation of motion 
\[  \ptl_{\nu}^2 b_{\mu}^A-\ptl_{\mu}\ptl_{\nu}b_{\nu}^A=0.  \]
Except for a string singularity, it is known that the configuration 
\[ \ct_{\mu}=\frac{n}{g}(\cos \theta -a)\ptl_{\mu}\phi,  \]
satisfies this equation, where $n$ is an integer, and $(r, \theta, \phi)$ are spherical coordinates.  
If $a=1$ ($a=-1$), the Dirac string appears on the negative (positive) $z$-axis.  
In Sect. 6, we choose $a=1$ as a concrete example, i.e., 
\begin{equation}
     \ct_{\mu}=\frac{n}{g}(\cos \theta -1)\ptl_{\mu}\phi=\frac{n}{g}\frac{z-r}{r(x^2+y^2)}(0,-y,x,0).   \lb{401}
\end{equation}
This monopole satisfies the equation of motion 
\[  \ptl_{\nu}(\ptl_{\nu} \ct_{\mu}-\ptl_{\mu}\ct_{\nu})=\epsilon_{\nu\mu\alpha\beta}\ptl_{\nu}\frac{n_{\alpha}}{n_{\rho} \ptl_{\rho}}k_{\beta}, \]
where
\[ n_{\alpha}=\delta_{\alpha 3},\quad k_{\beta}=\delta_{\beta 0}\frac{4\pi n}{g} \delta(x)\delta(y)\delta(z).  \]
We call $\ct_{\mu}$ magnetic potential.  

     In Sect. 3, we have shown the VEV $\Phi_4=\langle A_{\mu}^+A_{\mu}^-\rangle$ removes the tachyonic mass for 
the gluon.  What happens for the classical solution $b_{\mu}^A$?  If we substitute $A_{\mu}^A=a_{\mu}^A+b_{\mu}^A$ 
into Eq.(\ref{305}), Eq.(\ref{312}) changes to 
\begin{equation}
     V(\Phi_4) + (g^2\Phi_4 -m_4^2)[a_{\mu}^+a_{\mu}^- + (a_{\mu}^3)^2] + M_4^2a_{\mu}^+a_{\mu}^-
     +g^2\Phi_4[2b_{\mu}^3a_{\mu}^3 + (b_{\mu}^3)^2].  \lb{402} 
\end{equation}     
As $\Phi_4$ is given by Eq.(\ref{305}), the mass term for the magnetic potential appears as 
\begin{equation}
     g^2\Phi_4 (b_{\mu}^3)^2=m_4^2\ct_{\mu}\ct_{\mu}.  \lb{403}
\end{equation}
In the next section, we study the mass for $\ct_{\mu}$ by using the background covariant gauge.

\section{Massive magnetic potential}

     First we derive the Lagrangian with the field $\vp^A$ and the magnetic potential $\ct_{\mu}$ 
in the background covariant gauge \cite{ab}.
We devide $A_{\mu}^A$ into the classical part 
$b_{\mu}^A$ and the quantum fluctuation $a_{\mu}^A$ as 
\footnote{In this paper, as we are interested in the massive magnetic potential, 
we choose $b_{\mu}^A=\ct_{\mu}\delta^{A3}$.  
However, instead of $\ct_{\mu}$, we can introduce other classical solutions $B_{\mu}$ 
in the form $b_{\mu}^A=B_{\mu}\delta^{A3}$.} 
\[ A_{\mu}^A = b_{\mu}^A + a_{\mu}^A, \quad  b_{\mu}^A=\ct_{\mu}\delta^{A3}.  \]
Then the gauge transformation 
\begin{equation}
     \delta A_{\mu} = D_{\mu}(A) \varepsilon   \lb{501}
\end{equation}
is satisfied by 
\begin{equation}
     \delta a_{\mu} = D_{\mu}(b+a) \varepsilon,\  \delta b_{\mu} = 0,   \lb{502}
\end{equation}
where the notation $D_{\mu}(A)=(\ptl_{\mu}+gA_{\mu}\times)$ has been used.  
Choosing the gauge-fixing function 
\[ \tilde{G}(a)=D_{\mu}(b)a_{\mu}+\vp-w=(\ptl_{\mu}+gb_{\mu}\times)a_{\mu}+\vp-w,  \]
and applying the transformation (\ref{502}), 
we obtain the Lagrangian 
\begin{align}
 \cl&=  \cl_{inv}(b+a)+ \tilde{\cl}_{\vp}(a,b), \nonumber  \\
 \tilde{\cl}_{\vp}(a,b)&= -\frac{\alpha_1}{2}B^2 
 +B\cdot [D_{\mu}(b)a_{\mu}+\vp -w]+i\cb \cdot[D_{\mu}(b)D_{\mu}(b+a)+g\vp \times ]c+\frac{\vp^2}{2\alpha_2},  \lb{503}
\end{align}
where $w$ is a constant.  We require that $\vp^A$ belongs to the adjoint representation, and it transforms 
as $\delta \vp= g\vp\times \varepsilon$.  The last term $\vp^2/(2\alpha_2)$ is added because it is gauge invariant 
and necessary to yield the quartic ghost interaction term $-\alpha_2\bar{B}^2/2$ in Eq.(\ref{201}).  

     Since the ghost condensation happens by the quartic ghost interaction \cite{hs4}, the condensate $v=g\vp_0$ appears 
irrespective of $b_{\mu}$.  Thus, after the ghost condensation, $\tilde{\cl}_{\vp}(a,b)$ becomes 
\begin{equation}
     \tilde{\cl}_{\vp}(a,b)= -\frac{\alpha_1}{2}B^2 
 +B\cdot [D_{\mu}(b)a_{\mu}+\cvp ]+i\cb \cdot[D_{\mu}(b)D_{\mu}(b+a)+g(\vp_0+\cvp) \times ]c,   \lb{504}
\end{equation}
where $\vp(x)=\vp_0 + \cvp(x)$ and $w=\vp_0$.  We note Eq.(\ref{504}) is obtained from Eq.(\ref{203}) by replacing 
$\ptl_{\mu}$ to $D_{\mu}(b)$.  As we explained in Sect. 4, the VEV $g\langle \vp^A\rangle=v\delta^{A3}$ 
selects the unbroken U(1) direction, and the Abelian monopole $b_{\mu}^A=\ct_{\mu}\delta^{A3}$ 
is included in the Lagrangian consistently.

     Next we study mass terms.  Without $b_{\mu}$, a ghost loop brings about the tachyonic gluon 
mass term (\ref{301}).  We show that Eq.(\ref{301}) is obtained even if $b_{\mu}$ exists.  
To show it, we introduce the background gauge transformation 
\begin{equation}
    \delta b_{\mu} = D_{\mu}(b) \varepsilon, \quad \delta a_{\mu} = g a_{\mu} \times \varepsilon, \quad 
    \delta \cvp =g \cvp \times \varepsilon, \lb{505}
\end{equation}
which satisfies the transformation (\ref{501}).  When $v=0$, the Lagrangian $\cl_{inv}(b+a)$ and $\tilde{\cl}_{\vp}(a,b)$ 
in Eq.(\ref{504}) are invariant under the transformation (\ref{505}), if $\cb^A$, $c^A$ and $B^A$ transform 
in the adjoint representation.  
When $v\delta^{A3}\neq 0$, the ghost determinant 
\begin{equation}
     \int DcD\cb \exp[-\int dx i\cb \cdot(D_{\mu}(b)D_{\mu}(b+a)+v \times )c ]
 = \det [D_{\mu}(b)D_{\mu}(b+a)+v\times ]  \lb{506} 
\end{equation}
breaks the symmetry (\ref{505}).  However if we restrict Eq.(\ref{505}) to the U(1) 
transformation
\[ \delta b_{\mu}=-\frac{1}{g}\ptl_{\mu}\varepsilon , \quad \delta a_{\mu}^A=-\varepsilon f_{A3B}a_{\mu}^B,\quad|\varepsilon|\ll 1, \]
the determinant (\ref{506}) is invariant.  As tachyonic mass terms come from Eq.(\ref{506}), they must respect this 
U(1) symmetry.  So the terms $b_{\mu}^2$ and $b_{\mu}a_{\mu}^3$ are forbidden, and the terms 
$-L_1^2(a_{\mu}^a)^2$ and $-L_2^2(a_{\mu}^3)^2$ are allowed.  
We calculate the determinant (\ref{506}) in Appendix B, and obtain the tachyonic mass term 
\begin{equation}
 -m_4^2 [a_{\mu}^+a_{\mu}^- + (a_{\mu}^3)^2]  \lb{507} 
\end{equation}
as expected.  

     Finally, to remove the tachyonic mass, the procedure in the subsection 3.2 is applied.  
We replace $a_{\mu}^+a_{\mu}^-$ to $\Phi_4 + a_{\mu}^+a_{\mu}^-$, where $\Phi_4=\langle a_{\mu}^+a_{\mu}^- \rangle$, 
and add the source term $M_4^2a_{\mu}^+a_{\mu}^-$.  
As $\cl_{inv}(b+a)$ contains the term 
\begin{align}
 \frac{g^2}{4}\{f_{ABC}(b+a)_{\mu}^B(b+a)_{\nu}^C\}^2&=
\frac{g^2}{2}(a_{\mu}^+a_{\mu}^-)^2
+g^2(a_{\mu}^+a_{\mu}^-)\{(b_{\nu}^3+a_{\nu}^3)^2\} \nonumber \\
& - \frac{g^2}{2}(a_{\mu}^+)^2(a_{\nu}^-)^2
-g^2\{a_{\mu}^+(b_{\mu}^3+a_{\mu}^3)\}\{a_{\nu}^-(b_{\nu}^3+a_{\nu}^3)\},  \lb{508}
\end{align}
Eqs.(\ref{507}) and (\ref{508}) lead to 
\begin{equation}
 V(\Phi_4) + (g^2\Phi_4 -m_4^2)\{a_{\mu}^+a_{\mu}^- + (a_{\mu}^3)^2\} +M_4^2a_{\mu}^+a_{\mu}^-
     +g^2\Phi_4\{2b_{\mu}^3a_{\mu}^3 + (b_{\mu}^3)^2\}.  \lb{509}
\end{equation}
As $V(\Phi_4)$ takes the minimum value at $g^2\Phi_4=m_4^2$, Eq.(\ref{509}) 
becomes 
\[
  M_4^2a_{\mu}^+a_{\mu}^- + m_4^2\{2b_{\mu}^3a_{\mu}^3 + (b_{\mu}^3)^2\} 
\]
Namely, the magnetic potential $\ct_{\mu}$ has the mass term (\ref{403}).  

     Except for string singularity, the magnetic potential $\ct_{\mu}$ must be modified to satisfy 
the massive equation of motion 
\footnote{Eq.(\ref{401}) no longer satisfies Eq.(\ref{510}).  If we use a dual potential, a modified solution 
is obtained.  A concrete solution will be given in the next paper.}
\begin{equation}
   \left(\frac{\delta \cl}{\delta \ct_{\mu}}-\ptl_{\nu}\frac{\delta \cl}{\delta \ptl_{\nu}\ct_{\mu}}\right)_{a_{\rho}^A=0}
   = \ptl_{\nu}^2 \ct_{\mu}-\ptl_{\mu}\ptl_{\nu}\ct_{\nu}-2m_4^2\ct_{\nu}=0.  \lb{510}
\end{equation}
Then the term linear with respect to $a_{\mu}^A$ 
\[ a_{\mu}^3\left(\frac{\delta \cl}{\delta \ct_{\mu}}-\ptl_{\nu}\frac{\delta \cl}{\delta \ptl_{\nu}\ct_{\mu}}\right)_{a_{\rho}^A=0}  \]
vanishes.  Thus the final Lagrangian with massive magnetic potential becomes 
\begin{equation}
     \frac{1}{4}\left(\ptl_{\mu}\ct_{\nu}-\ptl_{\nu}\ct_{\mu}\right)^2 +m_4^2\ct_{\mu}\ct_{\mu}
      + \cl_{inv}(a)+ M_4^2a_{\mu}^+a_{\mu}^- + \cdots, \lb{511}
\end{equation}
where interaction terms with $a_{\mu}^A$ and $\ct_{\mu}$ are neglected.

\section{Massive non-Abelian magnetic potential}   

     The magnetic potential $\ct_{\mu}$ has string singularity.  We remove it by a singular 
gauge transformation \cite{afg}.  In this section, we use the matrix notation 
$D_{\mu}(b)=\ptl_{\mu} -ig b_{\mu}^AT_A$ with $(T_A)_{BC}=if_{BAC}$.  
We perform a singular gauge transformation for $A^A_{\mu}=b^A_{\mu}+a^A_{\mu}$ and $\vp^A=\vp_0 \delta^{A3}+\cvp^A$ as 
\begin{equation}
  A'^A_{\mu}T_A= U^{\dagger}A^A_{\mu}T_AU + \frac{i}{g}U^{\dagger}\ptl_{\mu}U\ ,\quad 
\vp'^AT_A = U^{\dagger}(\vp_0 \delta^{A3}+\cvp^A)T_AU.  \lb{601}
\end{equation}
The matrix $U$, which corresponds to the monopole $\ct_{\mu}$ in Eq.(\ref{401}), 
is $U=e^{-in\phi T_3}e^{i\theta T_2}e^{in\phi T_3}$.  As we explain in Appendix C, 
using this matrix $U$ and the monopole solution $b^A_{\mu}T_A=\ct_{\mu}T_3$, Eq.(\ref{601}) becomes 
\begin{equation}
     A'^A_{\mu}T_A= \left[a^1_{\mu}\hat{n}^A_1 +a^2_{\mu}\hat{n}^A_2+a^3_{\mu}\hat{n}^A
     -\frac{1}{g}(\hat{n}\times \ptl_{\mu}\hat{n})^A \right]T_A  \lb{602}
\end{equation}
and 
\begin{equation}
     \vp'^AT_A= \left(\vp_0\hat{n}^A + \cvp^B\hat{n}_B^A \right)T_A,  \lb{603}
\end{equation}
where $\hat{n}_1, \hat{n}_2$ and $\hat{n}_3=\hat{n}$ are defined in Eq.(\ref{c02}) with 
$\alpha=\theta$ and $\beta= -\gamma= n\phi$.  
We note the right hand side of Eq.(\ref{602}) is the gauge field decomposition by Cho \cite{cho1}, and  
Eq.(\ref{603}) shows the color vector $\hat{n}^A$ comes from the VEV $\vp_0\delta^{A3}$.  

     Now we regard the transformation (\ref{601}) as the background gauge transformation.  
The classical field has the inhomogeneous part as 
\begin{equation}
     b'_{\mu}=U^{\dagger}b_{\mu}U+\frac{i}{g}U^{\dagger}\ptl_{\mu}U, \quad a_{\mu}'=U^{\dagger}a_{\mu}U.  \lb{604}
\end{equation}
Then, as we show in Appendix C, Eqs.(\ref{602}) and (\ref{604}) indicate that the background gauge transformation 
is realized simply by the replacement 
\begin{equation}
     a_{\mu}^A \to  a_{\mu}^B\hat{n}_B^A, \quad
      \tilde{C}_{\mu}\delta^{3A} \to C_{\mu}^A=-\frac{1}{g}(\hat{n}\times \ptl_{\mu}\hat{n})^A .  \lb{605}
\end{equation}
We call $C_{\mu}^A$ non-Abelian magnetic potential.  
Let us give an example.  If we choose $n=1$, $\hat{n}$ and $C_{\mu}^A$ become 
\[ \hat{n}=\left(
\begin{array}{c}
 \sin \theta \cos \phi \\
 \sin \theta \sin \phi \\
 \cos \theta
\end{array}
\right), \quad 
C_{\mu}^A=\epsilon_{A\mu k}\frac{x^k}{gr^2}.  
\]
Although this configuration found by Wu and Yang \cite{wy} is singular at the origin, it has no string.

     Applying Eq.(\ref{605}) to the Lagrangian $\cl_{inv}(b+a)$ in Eq.(\ref{503}), we obtain 
\begin{align}
 \cl_{inv}=& \frac{1}{4}(F_{\mu\nu}+H_{\mu\nu})^2 +\frac{1}{4}(\hat{D}_{\mu}X_{\nu}-\hat{D}_{\nu}X_{\mu})^2   \nonumber\\
 &+\frac{g}{2}(F_{\mu\nu}+H_{\mu\nu})\hat{n}\cdot (X_{\mu}\times X_{\nu})+\frac{g^2}{4}(X_{\mu}\times X_{\nu})^2,  \lb{606}
\end{align}
where the following notation in Ref.~\cite{cho1} has been used: 
\begin{align}
 \hat{A}^A_{\mu}&=a^3_{\mu}\hat{n}^A + C_{\mu}^A, \quad  X^A_{\mu}=a^1_{\mu}\hat{n}^A_1 +a^2_{\mu}\hat{n}^A_2, \nonumber \\
 \hat{D}_{\mu}&= \ptl_{\mu} +g\hat{A}_{\mu}\times ,\quad F_{\mu\nu}=\ptl_{\mu}a^3_{\nu}-\ptl_{\nu}a^3_{\mu}, \nonumber \\
 H_{\mu\nu}\hat{n}^A&=\ptl{\mu}C_{\nu}^A-\ptl_{\nu}C_{\mu}^A+g(C_{\mu}\times C_{\nu})^A
 =-\frac{1}{g}(\ptl_{\mu}\hat{n}\times \ptl_{\nu}\hat{n})^A.  \lb{607}
\end{align}
Equation (\ref{606}) is the Lagrangian of the extended QCD \cite{cho1}.  
Next we apply the transformation (\ref{605}) to $\tilde{\cl}_{\vp}(a,b)$ in Eq.(\ref{504}) as well.  
However, as $\hat{n}_B\cdot\hat{n}_C=\delta_{BC}$, the tachyonic mass term (\ref{507}) is obtained again.  
Replacing $a_{\mu}^+a_{\mu}^-$ to $a_{\mu}^+a_{\mu}^- + \Phi_4$, and adding the source term 
$M_4^2a_{\mu}^+a_{\mu}^-$, the part that depends on $\Phi_4, m_4^2$ or 
$M_4^2$ becomes 
\begin{equation}
     V(\Phi_4) + (g^2\Phi_4 -m_4^2)[a_{\mu}^+a_{\mu}^- + (a_{\mu}^3)^2] + M_4^2a_{\mu}^+a_{\mu}^-
     +g^2\Phi_4[2C_{\mu}^Aa_{\mu}^3\hat{n}^A + (C_{\mu}^A)^2].   \lb{608} 
\end{equation}
Eq.(\ref{608}) is derived directly by applying the transformation (\ref{605}) to Eq.(\ref{509}).  
However there is a difference between Eqs.(\ref{509}) and (\ref{608}).  
Since $C_{\mu}^A\hat{n}^A=0$, the cross-term $g^2\Phi_4[2C_{\mu}^Aa_{\mu}^3\hat{n}^A]$ 
vanishes without using the equation of motion for $C_{\mu}^A$.  
Now we set $g^2\Phi_4=m_4^2$.  Then, from Eqs.(\ref{606}) and (\ref{608}), 
we obtain the Lagrangian 
\begin{equation}
   \frac{1}{4}(F_{\mu\nu}+H_{\mu\nu})^2 + m_4^2 (C_{\mu}^A)^2 +\frac{1}{4}(\hat{D}_{\mu}X_{\nu}-\hat{D}_{\nu}X_{\mu})^2 +
   \frac{M_4^2}{2}(X_{\mu}^A)^2 + \cdots, 
   \lb{609}
\end{equation}
where $2a_{\mu}^+a_{\mu}^-=(X_{\mu}^A)^2$ has been used.  
Eq.(\ref{609}) is the extended QCD with the massive non-Abelian magnetic potential 
and the massive off-diagonal gluons.

     We make a comment.  Neglecting the quantum fields, the Lagrangian ({\ref{609}) becomes 
\begin{equation}
   \cl_{SF}=\frac{1}{4}H_{\mu\nu}^2+m_4^2(C_{\mu}^A)^2=\frac{1}{4g^2}(\ptl_{\mu}\hat{n}\times \ptl_{\mu}\hat{n})^2
+\frac{m_4^2}{g^2}(\ptl_{\mu}\hat{n})^2.  \lb{610}
\end{equation}
This is the Skyrme-Faddeev Lagrangian \cite{fn}.  
The equation of motion for $\hat{n}$ comes from 
\[      Q^A(\hat{n})\delta \hat{n}^A=0,\quad
 Q^A(\hat{n})=\left( \frac{\delta \cl_{SF}}{\delta \hat{n}^A}-\ptl_{\mu}\frac{\delta \cl_{SF}}{\delta \ptl_{\mu}\hat{n}^A} \right).  \]
As $\delta \hat{n}^A$ satisfies $\hat{n}\cdot \delta \hat{n}=0$, this equation means that the component of 
$Q^A(\hat{n})$ that is perpendicular to $\hat{n}^A$ vanishes.  So, from Eq.(\ref{610}), we obtain 
\begin{equation}
     Q^A(\hat{n})= \frac{1}{g^2}\ptl_{\mu}[\ptl_{\nu}\hat{n}\times (\ptl_{\mu}\hat{n}\times \ptl_{\nu}\hat{n})]^A
+\frac{m_4^2}{g^2}\ptl_{\mu}^2\hat{n}^A,   \lb{611}
\end{equation}
and the perpendicular component gives \cite{cho1}
\begin{equation}
     -\frac{1}{g}\ptl_{\mu}(H_{\mu\nu})\ptl_{\nu}\hat{n}+\frac{m_4^2}{g^2}\hat{n}\times \ptl_{\mu}^2\hat{n} =0.  \lb{612}
\end{equation}

     Now we give an example.  The color vector $\hat{n}$ in Eq.(\ref{c07}) satisfies $\ptl_{\mu}(H_{\mu\nu})=0$.  
When $m_4^2\neq 0$, it must be modified to satisfy Eq.(\ref{612}).  
However we find it satisfies 
\[ 
\ptl_{\mu}^2\hat{n}=-\frac{2}{r^2}\hat{n}+\frac{1-n^2}{r^2\sin\theta}\left(
\begin{array}{c}
 \cos n\phi \\
 \sin n\phi \\
 0
\end{array}
\right).  
\]
So, if we choose $n=\pm 1$, the equations 
\[ \ptl_{\mu}(H_{\mu\nu})=0,\quad \hat{n}\times \ptl_{\mu}^2\hat{n}=0  \]
holds, and Eq.(\ref{612}) is fulfilled.  As $\int F_{\mu\nu}H_{\mu\nu} dx = -2\int a_{\nu}^3\ptl_{\mu}(H_{\mu\nu}) dx=0$, 
Eq.(\ref{609}) becomes 
\[   \frac{1}{4}H_{\mu\nu}^2 + m_4^2 (C_{\mu}^A)^2 + \frac{1}{4}F_{\mu\nu}^2 
+\frac{1}{4}(\hat{D}_{\mu}X_{\nu}-\hat{D}_{\nu}X_{\mu})^2 + \frac{M_4^2}{2}(X_{\mu}^A)^2 + \cdots.  \]

\section{Comment on the D=3 case}

     In D=3, the ghost condensation happens as well.  From Eq.(\ref{205}), the tachyonic gluon mass term is 
\[  -m_3^2 \left\{a_{\mu}^+a_{\mu}^- +\frac{\kappa_3}{2}(a_{\mu}^3)^2\right\},\quad
 m_3^2 = \frac{\alpha_2}{24\pi^2}g_3^4=\frac{\sqrt{2v_3}g_3^2}{12\pi},  \quad \kappa_3=\frac{3}{2}. \]
As in Sect. 5, we divide the gluon as $A_{\mu}^A=b_{\mu}^A+a_{\mu}^A$, add the source 
term $M_3^2 a_{\mu}^+a_{\mu}^-$, and introduce $\Phi_3$ as 
\[ a_{\mu}^+a_{\mu}^- \to \Phi_3 + a_{\mu}^+a_{\mu}^-, \quad \Phi_3=\langle a_{\mu}^+a_{\mu}^-\rangle.  \]
The terms that contain $\Phi_3, m_3^2$ or $M_3^2$ are 
\begin{equation}
 V(\Phi_3) + (g_3^2\Phi_3 -m_3^2+M_3^2)a_{\mu}^+a_{\mu}^- + \left(g_3^2\Phi_3 -\frac{\kappa_3}{2} m_3^2\right)(a_{\mu}^3)^2
     +g_3^2\Phi_3\{2b_{\mu}^3a_{\mu}^3 + (b_{\mu}^3)^2\},   \lb{701} 
\end{equation}
where 
\[
   V(\Phi_3)=\frac{g_3^2}{2}\Phi_3^2 -m_3^2\Phi_3,\quad   \Phi_3=\langle x|(-\Delta +M_3^2)^{-1}_{\mu\mu}|x \rangle. 
\]
The minimum of $V(\Phi_3)$ determines the VEV $\Phi_3$ and the mass $M_3$ as
\begin{equation}
 \Phi_3=\langle a_{\mu}^+a_{\mu}^- \rangle=\frac{m_3^2}{g_3^2}=\frac{\sqrt{2v_3}}{12\pi}, \quad
 \frac{m_3^2}{g_3^2}=\langle x|(-\Delta +M_3^2)^{-1}_{\mu\mu}|x \rangle.   \lb{702}
\end{equation}
Substituting Eq.(\ref{702}) into Eq.(\ref{701}), we find the mass terms 
\begin{equation}
    M_3^2a_{\mu}^+a_{\mu}^- +  \frac{m_3^2}{4}(a_{\mu}^3)^2+ m_3^2\{2b_{\mu}^3a_{\mu}^3 + (b_{\mu}^3)^2\}.  \lb{703}
\end{equation}
Namely, the components $a_{\mu}^{\pm}$ have the mass $M_3$, and the classical part $b_{\mu}^A$ acquires the mass $\sqrt{2}m_3$.  
However, different from the D=4 case, the component $a_{\mu}^3$ acquires the mass $m_3/\sqrt{2}=O(g_3^2)$.  
The value $\kappa_3=\frac{3}{2}\neq 2$ is crucial.  

     As the gauge field in $D=3$ (or $D=2+1$) is expected to have masses of this order \cite{na}.  
It is known that the $D=3+1$ gauge theory at high temperature becomes the effective $D=3$ gauge theory 
with the coupling constant $g_3=g\sqrt{T}$ \cite{ap}.  The mass of $O(g^2T)$, which is called 
the magnetic mass, is expected \cite{li}.  The mass terms $M_3^2a_{\mu}^+a_{\mu}^-$ and 
$m_3^2(a_{\mu}^3)^2/4$ may imply the existence of the magnetic mass at high $T$.

\section{Summary and comments}

     We have studied the SU(2) gauge theory in the nonlinear gauge.  In the 4-dimensional Euclidean space, 
the ghost condensation $\langle \vp^A \rangle =\vp_0\delta^{A3}\neq 0$ happens 
under the scale $\mu_0$, and the ghost loop yields the tachyonic gluon mass term 
$-m_4^2\{A_{\mu}^+A_{\mu}^- + (A_{\mu}^3)^2\}$.  We considered the effective potential for 
the LCO $A_{\mu}^+A_{\mu}^-$ up to $O(\hslash^2)$, 
and showed that $g^2\langle A_{\mu}^+A_{\mu}^-\rangle=m_4^2$ gives the minimum of the potential.  
The VEV $\Phi_4=\langle A_{\mu}^+A_{\mu}^-\rangle$ makes the diagonal gluon $A_{\mu}^3$ massless.  
In contrast, the off-diagonal gluons $A_{\mu}^{\pm}$ acquire the mass $M_4$ determined by Eq.(\ref{310}).  

     The field $\vp^A$ belongs to the adjoint representation of SU(2).  
Since the VEV $\vp_0\delta^{A3}$ selects the $A=3$ direction, the Abelian monopole 
solution $\ct_{\mu}$ can be introduced in this direction.  Combining the VEV 
$\langle A_{\mu}^+A_{\mu}^-\rangle$ and $\ct_{\mu}$, we find the magnetic potential $\ct_{\mu}$ becomes massive.  

     To remove the string singularity of the monopole solution $\ct_{\mu}$, we performed the singular gauge transformation 
in the background covariant gauge.  As $\vp_0\delta^{A3}$ transforms into $\vp_0 \hat{n}^A$, the field 
$\hat{n}^A$ which specifies the color direction appears, and the extended QCD \cite{cho1} is derived.  
In addition, the non-Abelian magnetic potential $C_{\mu}^A$ transformed from $\ct_{\mu}\delta^{A3}$ 
becomes massive.  If quantum fields are neglected, the Skyrme-Faddeev Lagrangian is obtained.  

     In the 3-dimentional Euclidean space, the phenomena similar to the D=4 case happen.  However, 
different from the D=4 case, the diagonal component $a_{\mu}^3$ becomes massive.  

     We make some comments.  
\begin{enumerate}
 \item[(1)] In the maximally Abelian gauge, the lattice simulation shows that 
the off-diagonal gluons are massive and the diagonal gluons are nearly massless \cite{sug}.  
In the present model, the VEV $\vp_0$ breaks the global SU(2) symmetry to U(1).  
So the result that $A_{\mu}^{\pm}$ and $A_{\mu}^3$ have different masses may be reasonable.  
In addition, the masslessness of the component $A_{\mu}^3$ may be related to the remaining U(1) symmetry.  
 \item[(2)] Since magnetic potential $\tilde{C}_{\mu}$ couples with the off-diagonal components $A_{\mu}^{\pm}$, 
$\tilde{C}_{\mu}$ can become 
massive by the VEV $\langle A_{\mu}^+A_{\mu}^-\rangle$.  Namely we can obtain massive 
magnetic potential without monopole condensation.  
 \item[(3)]  To show the quark confinement, the Abelian dominance is often assumed \cite{ei,suz}.  
In this assumption, the off-diagonal components $A_{\mu}^{\pm}$ are neglected.  Our result insists 
that these components play an important role to 
realize the Abelian dominance.  
\footnote{The components $\rho d\hat{n}$ and $\sigma d\hat{n}\times \hat{n}$ 
in Ref.~\cite{fn} correspond to $A_{\mu}^{\pm}$ in this paper.  Condensations related to 
$\rho$ and $\sigma$ are discussed in Ref.~\cite{fn}.}  
 \item[(4)]  In Ref.~\cite{cdg}, based on the LCO formalism in Ref.~\cite{vds}, the LCO 
$(\bar{c}\times c)^A$ is studied in the Landau gauge, together with the LCO $(A_{\mu}^A)^2=(A_{\mu}^a)^2+(A_{\mu}^3)^2$.  
This formalism introduces sources $\omega^A$ and 
$J$ for $\bar{c}\times c$ and $(A_{\mu}^A)^2$ respectively in a BRS-exact form.  From the renormalizability, 
the terms $(\omega^A)^2$ and $J^2$ are necessary.  To delete 
these terms, the Hubbard-Stratonovich transformation is applied, and the fields $\phi^A$ and $\sigma$, which correspond to 
$\bar{c}\times c$ and $(A_{\mu}^A)^2$ respectively, are introduced.  
The final Lagrangian they used is $\mathcal{L}=\mathcal{L}_{YM}+\mathcal{L}_{GF}+\mathcal{L}_{ext}$, where 
\begin{align*}
  \mathcal{L}_{YM}&=\frac{1}{4}(F_{\mu\nu}^A)^2, \quad \mathcal{L}_{GF}=B^A(\partial_{\mu}A_{\mu}^A), \\
  \mathcal{L}_{ext}&=\mathcal{L}_{ext}(\phi) + \mathcal{L}_{ext}(\sigma),  \\
  \mathcal{L}_{ext}(\phi)&=\frac{(\phi^A)^2}{2g^2\rho}+\frac{1}{\rho}\phi^A(\bar{c}\times c)^A
  +\frac{g^2}{2\rho}(\bar{c}\times c)^2 - \omega^A\frac{\phi^A}{g},  \\
  \mathcal{L}_{ext}(\sigma)&=\frac{\sigma}{2g^2\zeta}+\frac{\sigma}{2g\zeta}(A_{\mu}^A)^2 
  +\frac{1}{8\zeta}(A_{\mu}^AA_{\mu}^A)^2 - J\frac{\sigma}{g}. 
\end{align*}
The one-loop effective potential $V(\phi)+V(\sigma)$ was calculated by using the free propagators derived from 
$\mathcal{L}_{YM}+\mathcal{L}_{GF}$ and the interactions in $\mathcal{L}_{ext}$.  
Then the conditions $\frac{dV(\phi)}{d\phi}=0$ and $\frac{dV(\sigma)}{d\sigma}=0$ give 
the VEVs $\langle \phi^3 \rangle=-g^2\langle (\bar{c}\times c)^3 \rangle$ and 
$\displaystyle \langle \sigma \rangle=-\frac{g}{2}\langle (A_{\mu}^A)^2 \rangle$, respectively.  

     When the VEV $\langle (\bar{c}\times c)^3 \rangle$ exists, ghost loop yields the tachyonic gluon masses.  
As in Sect. 2, the magnitude for $A_{\mu}^3$ is twice that for $A_{\mu}^{\pm}$.  In the presence of the VEV 
$\langle (A_{\mu}^A)^2 \rangle$, gluons acquire a mass.  The latter mass compensates the former 
tachyonic masses.  Thus gluons have the usual masses, and the mass for $A_{\mu}^3$ is different from that for 
$A_{\mu}^{\pm}$.  

     In their approach, the interactions which yield these VEVs 
are introduced irrelevant to the original Lagrangian 
$\mathcal{L}_{YM}+\mathcal{L}_{GF}$.  As a result, the VEV $\langle (A_{\mu}^A)^2 \rangle$ appears 
irrespective of the VEV $\langle (\bar{c}\times c)^3 \rangle$.

 \item[(5)]  In our procedure, we need the action in the nonlinear gauge.  The quartic ghost interaction yields 
the VEV $\langle (\bar{c}\times c)^3 \rangle$, 
and ghost loop brings about the tachyonic gluon masses.  The VEV 
$\langle A_{\mu}^+A_{\mu}^-\rangle=\frac{1}{2}\langle A_{\mu}^aA_{\mu}^a\rangle$ appears to eliminate the 
tachyonic mass for $A_{\mu}^{\pm}$.

     In the Landau gauge, there is no quartic ghost interaction.  However, in the low energy region, 
gauge field configurations on the Gribov horizon contribute to the partition function $Z$.  These configurations 
give rise to zero-modes of the ghost operator $\partial_{\mu}D_{\mu}$, and make $Z$ vanishes.   

     However, these zero-modes can produce the quartic ghost interaction \cite{hs3}.  The Landau gauge 
changes automatically to the nonlinear gauge.  Thus, even if we start from the Landau gauge, 
effective quartic ghost interaction is produced, and 
we can apply the scenario proposed in this article.

     We note the scale $\mu_0$ in Eq.(\ref{204}) depends on the parameter $\alpha_2$ at the cut-off scale $\Lambda$.  
One-loop calculation shows that $\alpha_2$ has the ultraviolet fixed point $\alpha_2=\beta_0/2$, 
where $\beta_0$ is the coefficient of the beta function $\displaystyle \beta=-\frac{\beta_0}{16\pi^2}g^3$.  
By substituting $\alpha_2=\beta_0/2$ into $\mu_0$, 
we find $\mu_0$ coincides with the QCD scale parameter $\Lambda_{\mathrm QCD}$\cite{hs2}.  
\end{enumerate}

\appendix

\section{$\langle (A_{\mu}^3)^2\rangle=0$}

    To eliminate the tachyonic mass terms
\begin{equation}
     -m_D^2A_{\mu}^+A_{\mu}^- -m_D^2 \frac{\kappa_D}{2}(A_{\mu}^3)^2, \lb{a01}
\end{equation}
the VEV $\langle A_{\mu}^+A_{\mu}^-\rangle$ is necessary.  
In this Appendix, we show, contrary to $\langle A_{\mu}^+A_{\mu}^-\rangle$, the VEV $\langle (A_{\mu}^3)^2\rangle$ vanishes.  

    To apply the procedure in subsection 3.2, we substitute  
\begin{equation}
     A_{\mu}^+A_{\mu}^- \to \Phi_D + A_{\mu}^+A_{\mu}^-,\ \Phi_D=\langle A_{\mu}^+A_{\mu}^-\rangle, \quad  
     (A_{\mu}^3)^2 \to \Psi_D + (A_{\mu}^3)^2,\ \Psi_D=\langle (A_{\mu}^3)^2\rangle  \lb{a02}
\end{equation}
into Eqs.(\ref{305}) and (\ref{a01}).  Then the part that contains $\Phi_D$ or $\Psi_D$ is given by 
\[
 V(\Phi_D,\Psi_D) = V_1+V_2,\quad V_1=\frac{g^2}{2}\Phi_D^2 - m_D^2\Phi_D,\quad 
 V_2= \left(g^2\Phi_D - m_D^2\frac{\kappa_D}{2}\right)\Psi_D. 
\] 
If we set $g^2\Phi_D=\lambda m_D^2$, we obtain 
\begin{equation}
     V(\Phi_D,\Psi_D) = V_1(\lambda)+V_2(\lambda), \quad V_1(\lambda)=\frac{m_D^4}{2g^2}\lambda(\lambda-2), \quad
      V_2(\lambda)= m_D^2\left(\lambda - \frac{\kappa_D}{2}\right)\Psi_D.  \lb{a03}
\end{equation}
The potential $V_1$ has the minimum value $\displaystyle -\frac{m_D^4}{2g^2}$ at $\lambda=1$.  
As $\Psi_D \geq 0$, and $V_2$ is linear with respect to $\Psi_D$, the coefficient must satisfy 
$(g^2\Phi_D - \kappa_D m_D^2/2)\geq 0$.  
Thus we find 
\begin{equation}
     V_2(\lambda)=m_D^2\left(\lambda - \frac{\kappa_D}{2}\right)\Psi_D \geq 0.  \lb{a04}
\end{equation}
Since the local minimum condition $\displaystyle \frac{\delta V}{\delta \Phi_D}=0$ leads to 
\begin{equation}
     g^2\Phi_D+g^2\Psi_D-m_D^2 = (\lambda -1)m_D^2 +g^2\Psi_D=0,  \lb{a05}
\end{equation}
from Eqs.(\ref{a04}) and (\ref{a05}), we obtain 
\[
    \frac{\kappa_D}{2}\leq \lambda \leq 1.  
\]
Now we show $\Psi_D=0$.  When $D=4$, as $\kappa_4=2$,  $V_1$ and $V_2$ have their minimum values at $\lambda=1$, 
and Eq.(\ref{a05}) leads to $\Psi_4=0$.  When $D=3$, there are two possibilities 
to minimize $V_2$.  As $\kappa_3=3/2$, one way is to choose $\lambda=3/4$, where  has been used.  
However since $V_1(\lambda=3/4)>V_1(\lambda=1)$, $V_1$ does not have minimum value.  Another way 
is to set $\Psi_3=0$.  Then Eq.(\ref{a05}) gives $\lambda=1$, and $V_1$ is minimized.

     We make a comment.  Historically, the VEV $\langle (A_{\mu}^A)^2\rangle$ was 
discussed in the framework of the operator product expansion (OPE) \cite{ls, by}, 
and the relation with monopoles was considered \cite{gz}.  
When $k_{\mu} \to \infty$, the OPE gives \cite{by} 
\[ \langle T(A_{\nu}^B(-k)A_{\rho}^C(k))\rangle = (c_0)_{\nu\rho}^{BC}(k) + 
(c_2)_{\nu\rho}^{BC}(k)\langle :(A_{\mu}^A)^2:\rangle + \cdots.  \]
However the VEV $\langle :(A_{\mu}^A)^2:\rangle $ is different from 
$\Phi_D=\langle A_{\mu}^+A_{\mu}^-\rangle$, because $\Phi_D$ appears as $k_{\mu} \to 0$, and 
the component $A_{\mu}^3$ does not condense.  
To make $A_{\mu}^{\pm}$ massive, the VEV $\langle A_{\mu}^+A_{\mu}^-\rangle$ and $\langle \cb^a c^a\rangle$ 
was considered in Ref.~\cite{kon2}.  In our approach, although $A_{\mu}^{\pm}$ becomes massive, 
the significant role of $\Phi_D$ is to remove the tachyonic masses.

\section{Derivation of Eq.(\ref{507})}

     We calculate the ghost determinant Eq.(\ref{506}) given by  
\begin{align*}
 & \det[D_{\mu}(b)D_{\mu}(A)+v \times] = \exp[-\int dx V_{gh}],\\
 & V_{gh} = -\mathrm{tr}\int\frac{d^Dk}{(2\pi)^D} \ln[D_{\mu}(b)D_{\mu}(A)+v\times ],  
\end{align*}
where the classical part is $b_{\mu}^A=b_{\mu}^3\delta^{A3}$, and $A_{\mu}^A=b_{\mu}^A+a_{\mu}^A$.  
Up to the second order of the fields, we find 
\begin{align}
   V_{gh}=&-\int\frac{d^Dk}{(2\pi)^D}\ln [k^4 + v^2 +2vigk_{\mu}(A_{\mu}^3+b_{\mu}^3) \nonumber \\
       &-g^2k_{\mu}k_{\nu}\{A_{\mu}^aA_{\nu}^a + (A_{\mu}^3+b_{\mu}^3)(A_{\nu}^3+b_{\nu}^3)\}+
       2g^2k^2A_{\mu}^3b_{\mu}^3+\cdots ]  \lb{b01}
\end{align}
The term $-g^2k_{\mu}k_{\nu}\{A_{\mu}^aA_{\nu}^a +(A_{\mu}^3+b_{\mu}^3)(A_{\nu}^3+b_{\nu}^3)\}$ in Eq.(\ref{b01}) gives 
\[ -\int \frac{d^Dk}{(2\pi)^D} \frac{-g^2k_{\mu}k_{\nu}}{k^4+v^2} \{A_{\mu}^aA_{\nu}^a +(A_{\mu}^3+b_{\mu}^3)(A_{\nu}^3+b_{\nu}^3)\} \]
Using 
\[ \int \frac{d^4k}{(2\pi)^4} \frac{k_{\mu}k_{\nu}}{k^4+v^2} = -\frac{\delta_{\mu\nu}}{2}\frac{v}{64\pi}, \quad 
\int \frac{d^4k}{(2\pi)^4} \frac{k_{\mu}k_{\nu}}{(k^4+v^2)^2} = \frac{\delta_{\mu\nu}}{4}\frac{v}{64\pi}\]
we obtain \cite{hs1}
\[
  V_{gh(1)}= \frac{1}{2}\left(-\frac{g^2v}{64\pi}\right)\{(A_{\mu}^a)^2 +(A_{\mu}^3+b_{\mu}^3)^2\}
\]
In the same way, the term $2vigk_{\mu}(A_{\mu}^3+b_{\mu}^3)$ in Eq.(\ref{b01}) gives 
\[ V_{gh(2)}=-\int \frac{d^Dk}{(2\pi)^D} \frac{2g^2v^2k_{\mu}k_{\nu}}{(k^4+v^2)^2} (A_{\mu}^3+b_{\mu}^3)(A_{\nu}^3+b_{\nu}^3) 
= \frac{1}{2}\left(-\frac{g^2v}{64\pi}\right)(A_{\mu}^3+b_{\mu}^3)^2,  \]
and the term $2k^2g^2A_{\mu}^3b_{\mu}^3$ in Eq.(\ref{b01}) gives 
\[ V_{gh(3)}=-\int \frac{d^Dk}{(2\pi)^D} \frac{2g^2k^2}{k^4+v^2}A_{\mu}^3b_{\mu}^3=\frac{4g^2v}{64\pi}A_{\mu}^3b_{\mu}^3. \]
Thus, substituting $A_{\mu}^a=a_{\mu}^a$ and $A_{\mu}^3=b_{\mu}^3+a_{\mu}^3$, we obtain 
\[ V_{gh}=\sum_{n=1}^3 V_{gh(n)}= \frac{1}{2}\left(-\frac{g^2v}{64\pi}\right)\{(a_{\mu}^a)^2 +2(a_{\mu}^3)^2\} \]

\section{Singular gauge transformation}

     Following Ref.\cite{cho}, we consider the gauge transformation with the matrix 
\[ U=e^{i\gamma T_3}e^{i\alpha T_2}e^{i\beta T_3} \]
where $(T_B)_{AC}=i f_{ABC}$.  This matrix $U$ satisfies 
\begin{equation}
 \frac{\langle \vp\rangle}{\vp_0}=\begin{pmatrix}0\cr 0\cr 1\cr \end{pmatrix}=U \hat{n},\quad 
   \hat{n}=\begin{pmatrix}\sin\alpha \cos\beta \cr
                      \sin\alpha \sin\beta \cr
                      \cos\alpha \cr \end{pmatrix}.    \lb{c01}
\end{equation}
and $U^{\dagger}T_AU=T_B\hat{n}^B_A$, where 
\begin{equation}
\hat{n}_1=\begin{pmatrix}\cos\alpha \cos\beta \cos\gamma-\sin\beta\sin\gamma \cr
                      \cos\alpha \sin\beta \cos\gamma+\cos\beta\sin\gamma \cr
                      -\sin\alpha \cos\gamma \cr\end{pmatrix},
   \hat{n}_2=\begin{pmatrix}-\cos\alpha \cos\beta \sin\gamma-\sin\beta\cos\gamma \cr
                      -\cos\alpha \sin\beta \sin\gamma+\cos\beta\cos\gamma \cr
                      \sin\alpha \sin\gamma \cr \end{pmatrix},
   \hat{n}_3=\hat{n}.  \lb{c02}
\end{equation}
These color vectors satisfy the orthonormality  
\[ \hat{n}_A^C\hat{n}_B^C=\delta_{AB}.  \]
Under this transformation, the gauge field transforms as 
\begin{align}
 A'^A_{\mu}T_A&= U^{\dagger}A^A_{\mu}T_AU + \frac{i}{g}U^{\dagger}\ptl_{\mu}U \nonumber \\
 &= \left[A^1_{\mu}\hat{n}^A_1 +A^2_{\mu}\hat{n}^A_2+A^3_{\mu}\hat{n}^A-\frac{1}{g}\left\{(\hat{n}\times \ptl_{\mu}\hat{n})^A + 
 (\cos \alpha \ptl_{\mu}\beta+\ptl_{\mu}\gamma)\hat{n}^A\right\}\right]T_A.    \lb{c03}
\end{align}
If we write 
\begin{equation}
     A^A_{\mu}=b^A_{\mu}+a^A_{\mu}, \quad b_{\mu}^A=\frac{1}{g}(\cos \alpha \ptl_{\mu}\beta+\ptl_{\mu}\gamma)\delta^{A3}, \lb{c04}
\end{equation}
Eq.(\ref{c03}) becomes 
\begin{equation}
     A'^A_{\mu}T_A= \left[a^1_{\mu}\hat{n}^A_1 +a^2_{\mu}\hat{n}^A_2+a^3_{\mu}\hat{n}^A-
     \frac{1}{g}\left\{(\hat{n}\times \ptl_{\mu}\hat{n})^A \right\}\right]T_A.  \lb{c05}
\end{equation}
If we regard the transformation (\ref{c05}) as the background gauge transformation 
\[ U^{\dagger}b_{\mu}U+\frac{i}{g}U^{\dagger}\ptl_{\mu}U,\quad U^{\dagger}a_{\mu}U, \]
Eq.(\ref{c05}) implies that $a_{\mu}^A$ and $b_{\mu}^A$ transform as 
\begin{equation}
   a_{\mu}^A \to  a_{\mu}^B\hat{n}_B^A, \quad b^A_{\mu} \to -\frac{1}{g}(\hat{n}\times \ptl_{\mu}\hat{n})^A .  \lb{c06}
\end{equation}

     Now, using the spherical coordinates $(r,\theta,\phi)$ and integer $n$, we set the angles 
as $\alpha=\theta, \beta= -\gamma=n \phi$.  
Then $b_{\mu}^A$ and $\hat{n}^A$ become 
\begin{equation}
     b_{\mu}^A=\ct_{\mu}\delta^{3A},\quad 
     \hat{n}=\left(
\begin{array}{c}
 \sin \theta \cos n\phi \\
 \sin \theta \sin n\phi \\
 \cos \theta
\end{array}
\right),  \lb{c07}
\end{equation}
where $\ct_{\mu}$ is the Abelian monopole in Eq.(\ref{401}).  
The corresponding non-Abelian monopole is 
\begin{equation}
  C_{\mu}^A=-\frac{1}{g}(\hat{n}\times \ptl_{\mu}\hat{n})^A
   =\frac{1}{g}\left(
\begin{array}{c}
 \sin n\phi \ptl_{\mu}\theta + \sin\theta \cos \theta \cos n\phi \ptl_{\mu}(n\phi)  \\
 -\cos n\phi \ptl_{\mu}\theta + \sin \theta \cos\theta \sin n\phi \ptl_{\mu}(n\phi) \\
 -\sin^2 \theta \ptl_{\mu}(n\phi)
\end{array}
\right).  \lb{c08}
\end{equation}
The field strength $H_{\mu\nu}$ is defined in Eq.(\ref{607}).  If we use Eq.(\ref{c08}), it becomes 
\[ H_{\mu\nu}=-\frac{1}{g}\sin\theta \{\ptl_{\mu}\theta \ptl_{\nu}(n\phi)-\ptl_{\nu}\theta \ptl_{\mu}(n\phi)\}, \]
and it satisfies $\ptl_{\mu}H_{\mu\nu}=0$.  In the same way, $\hat{n}$ in Eq.(\ref{c07}) satisfies 
\[ \ptl_{\mu}^2\hat{n}=-\frac{2}{r^2}\hat{n}+\frac{1-n^2}{r^2\sin\theta}\left(
\begin{array}{c}
 \cos n\phi \\
 \sin n\phi \\
 0
\end{array}
\right).  \]

\section{BRS symmetry and global gauge symmetry}

     In this Appendix, for the sake of explanation, we use the operator formalism \cite{ko}.  
The BRS transformation is $\delta_B$ and the BRS charge is $Q_B$.  A state $|\ph \rangle$ in the physical subspace 
satisfies the condition $Q_B |\ph \rangle=0$.  

\subsection{BRS symmetry}

     First we show that the constant $w$ in Eq.(\ref{202}) must be chosen as $w^A=\vp_0 \delta^{A3}$ to 
preserve the BRS symmetry.  
The Lagrangian (\ref{202}) is invariant under the BRS transformation 
\[
 \delta_B A_{\mu}=D_{\mu}c,\ \delta_B c= -\frac{g}{2}c\times c,\ \delta_B \cb =iB,\ 
 \delta_B \vp = g\vp\times c,\ \delta_B w=0.  
\]
Using the equation of motion for $B^A$, we obtain 
\begin{equation}
  \langle 0|\alpha_1 B^A |0 \rangle =  \vp_0\delta^{A3} -w^A, \lb{d01}
\end{equation}
where $\langle 0| A_{\mu}|0 \rangle =0$ and $\langle 0| \vp^A|0 \rangle =\vp_0 \delta^{A3}$ 
have been used.  As $B=-i\delta_B \cb$, we must set $w^A=\vp_0 \delta^{A3}$ to obtain 
$\langle 0| \delta_B \cb|0 \rangle =\langle 0|\{iQ_B, \cb\}|0 \rangle =0$.  

     Next we consider the VEV $\langle 0| A_{\mu}^+A_{\mu}^-|0 \rangle$.  
Since the operator $A_{\mu}^+A_{\mu}^-$ satisfies 
\begin{equation}
   \delta_B (A_{\mu}^+A_{\mu}^-) = A_{\mu}^a(D_{\mu}c)^a\neq 0 ,  \lb{d02}
\end{equation}
it is not BRS-invariant.  If there is an operator $\Omega$ which satisfies 
$\delta_B \Omega = A_{\mu}^+A_{\mu}^-$, 
the BRS symmetry is broken spontaneously by the VEV $\langle 0| A_{\mu}^+A_{\mu}^-|0 \rangle$.  However, 
as $\delta_B^2=0$, Eq,(\ref{d02}) implies such an operator $\Omega$ does not exist.  
\footnote{In the case of $\vp^A$, the equation of motion for $B^A$ gives Eq.(\ref{d01}).  However, in the 
case of $\langle 0| A_{\mu}^+A_{\mu}^-|0 \rangle$, any colored fields don't give the equation of motion 
with $(A_{\mu}^a)^2$.}
So the VEV $\langle 0|A_{\mu}^+A_{\mu}^-|0\rangle \neq 0$ does not contradict the BRS invariance of the vacuum.  

     We make two comments.  First, since there is no ghost number violating interaction nor such a condensate, 
the ghost number should be conserved.  Therefore, although $\delta_B A_{\mu}^+A_{\mu}^-\neq 0$, its VEV satisfies 
\[ \delta_B \langle 0|A_{\mu}^+A_{\mu}^-|0\rangle = \langle 0|A_{\mu}^a(D_{\mu}c)^a|0\rangle = 0. \]
Second, the anti-BRS symmetry is broken in this model \cite{hs3}.  However the unitarity of the model is guaranteed 
by the BRS symmetry.

\subsection{Global SU(2) symmetry}

          Using the constant small parameter $\theta$, the global color transformation is defined by 
$\dl_{\theta} \Sigma = \theta \times \Sigma$, where $\Sigma$ represents all the fields in $\cl_{\vp}$.  
Since $\langle 0|\dl_{\theta}\vp^A|0 \rangle =f^{AB3}\theta^B\vp_0$, this symmetry breaks down 
spontaneously to U(1), and the fields $\vp^{\pm}$ are Goldstone bosons.  In addition, 
as $\dl_{\theta} \cl_{\vp}=-w\cdot ( \theta \times B)$, the nonzero constant $w$ breaks 
this symmetry at the Lagrangian level.  However, as $B=-i\dl_B\cb$, this breaking term 
$\dl_{\theta}\cl_{\vp}=-i(w\times \theta)\cdot \dl_B \cb$ is BRS-exact.  
Since physical states satisfy $Q_B |\ph \rangle=0$, we find 
\[
    \langle \ph_2|\dl_{\theta}\cl_{\vp}|\ph_1 \rangle=(w\times \theta)\cdot \langle \ph_2|\{Q_B, \cb\}|\ph_1 \rangle=0.  
\]
Thus the breaking 
term does not contribute to amplitudes between physical states \cite{hs3}.

\end{document}